\documentclass{iau} 
\usepackage{graphicx}
\usepackage[english]{babel}
\usepackage[utf8]{inputenc}
\usepackage[T1]{fontenc}
\usepackage{setspace}
\usepackage{multicol}
\usepackage{xcolor}
\usepackage{wrapfig,lipsum}
\usepackage{epsfig}
\usepackage{amssymb}
\usepackage{blindtext}
\usepackage{hyperref}

\title[Variability of masers and MYSOs] 
{Infrared variability, maser activity, and accretion of massive young stellar objects}

\author[Bringfried Stecklum \& Alessio Caratti o Garatti]   
{Bringfried Stecklum$^1$, 
 Alessio Caratti o Garatti$^2$, Klaus Hodapp$^3$, Hendrik Linz$^4$,
Luca Moscadelli$^5$, and Alberto Sanna$^6$
}

\affiliation{$^1$Thüringer Landessternwarte Tautenburg, \\ Sternwarte 5,
D-07778 Tautenburg, Germany \\ email: {\tt stecklum@tls-tautenburg.de} \\[\affilskip]
$^2$Dublin Institute for Advanced Studies, Dublin, Ireland\\[\affilskip]
$^3$Institute for Astronomy, Hilo, USA\\[\affilskip]
$^4$Max-Planck Institut für Astronomie, Heidelberg, Germany\\[\affilskip]
$^5$INAF, Firenze, Italy\\[\affilskip]
$^6$Max-Planck Institut für Radioastronomie, Bonn, Germany
}

\pubyear{2017}
\volume{336}  
\jname{Astrophysical Masers: Unlocking the Mysteries of the Universe}
\editors{A. Tarchi, M.J. Reid \& P. Castangia, eds.}
\begin{document}

\maketitle

\begin{abstract}
Methanol and water masers indicate young stellar objects. They often exhibit flares, and a fraction shows periodic activity. Several mechanisms might explain this behavior but the lack of concurrent infrared (IR) data complicates to identify the cause. Recently, 6.7 GHz methanol maser flares were observed, triggered by accretion bursts of high-mass YSOs which confirmed the IR-pumping of these masers. This suggests that regular IR changes might lead to maser periodicity. Hence, we scrutinized space-based IR imaging of YSOs associated with periodic methanol masers. We succeeded to extract the IR light curve from NEOWISE data for the intermediate mass YSO G107.298+5.639. Thus, for the first time a relationship between the maser and IR variability could be established. While the IR light curve shows the same period of $\sim$34.6 days as the masers, its shape is distinct from that of the maser flares. Possible reasons for the IR periodicity are discussed.
\keywords{masers, molecules, dust, stars: formation, individual  
\href{http://simbad.u-strasbg.fr/simbad/sim-id?Ident=[TGJ91]+S255+NIRS+3&NbIdent=1&Radius=2&Radius.unit=arcmin&submit=submit+id}{([TGJ91] S255 NIRS 3}, \href{http://simbad.u-strasbg.fr/simbad/sim-id?Ident=[PFG2013]+I22198-MM2-S&NbIdent=1&Radius=2&Radius.unit=arcmin&submit=submit+id}{[PFG2013] I22198-MM2-S})}
\end{abstract}

\firstsection 
\section{Introduction}
Class II methanol masers are signs of luminous young stellar objects (YSOs) (\cite{breen_confirmation_2013}). The strong infrared (IR) radiation from high-mass YSOs  (HMYSOs) heats up dust in their immediate environment which causes molecules, originally frozen onto the grains, to sublimate. Both, the high-column density of molecules in the gas phase  as well as the strong mid-IR radiation due to the thermal dust emission are thought to be essential for the excitation of these masers (\cite{sobolev_pumping_1997}).
They often show flare activity,
and a few dozens of them vary periodically within $\sim$30\dots800 
days (\cite{goedhart_periodicity_2014,szymczak_discovery_2015}). 
In the absence of complementary data, in particular time-resolved IR photometry, this variability seemed to be enigmatic. Thus, it is not surprising that various models were brought up to explain periodic masers, which rely on vastly differing mechanisms, and have no constraints yet other than to reproduce the maser light curves. These comprise masers in the atmosphere of evaporating icy planets orbiting OB stars
(\cite{slysh_1998}),  modulation of the radio continuum seed radiation due to variable colliding binary winds  (\cite{van_der_walt_methanol_2011}) or eclipsing massive binaries (\cite{maswanganye_new_2015}),  variation of the IR pumping radiation due to periodic accretion from a circumbinary disk  (\cite{araya_quasi-periodic_2010}), protostellar pulsations at high accretion rates (\cite{inayoshi_direct_2013}), or  heating by accretion flow shocks in binary systems (\cite{parfen_sob_2014}). Recent conclusions from the statistics of both maser strength and variability point to a correlation between maser and MYSO luminosity while the maser variability appears to  be anti-correlated with the latter. 
This suggests that IR flux variations may drive maser flares (\cite{szymczak_stats_2017}). 

\section{Accretion bursts and IR pumping of Class II methanol masers}
Episodic accretion bursts are well known among low- and intermediate mass YSOs. They are caused when matter, piled-up at the inner circumstellar disk for various reasons, is being dumped onto the young star within rather a short time. For the first events found,  the resulting luminosity increase was discovered in the optical as brightness rise of up to 5 mags. They were classified into two categories, FUors and EXors, according to the properties of their prototype objects (\cite{herb77,herb89}). In the meanwhile, it turned out that these just represent the tip of the iceberg, seen in the latest stage of protostellar evolution. Recent IR surveys revealed that embedded YSOs are inherently variable, with the deeply embedded ones varying strongest (\cite{pena_2017a}). Spectroscopy of these variables confirmed their eruptive nature, showing a mixture of FUor/EXor features (\cite{pena_2017b}). If disk-mediated accretion is a pathway to form OB stars, episodic accretion should occur for their precursors, too. 
\begin{wrapfigure}{r}{0.45\textwidth}
\vspace*{-.45cm}
  \begin{center}
    \includegraphics[width=0.475\textwidth]{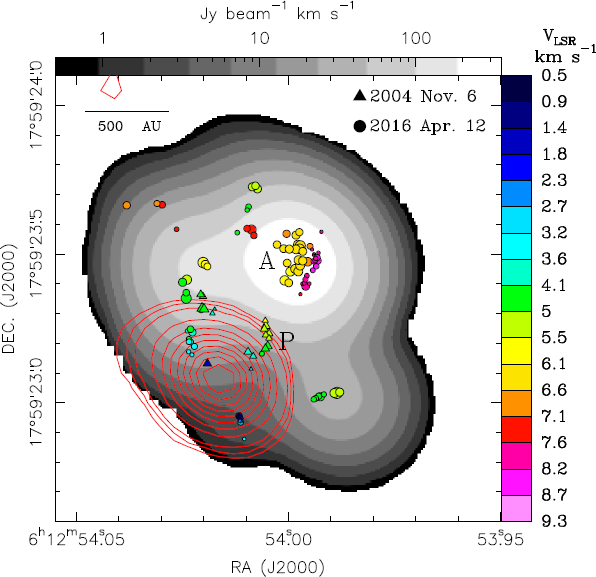}
  \end{center}
  \caption{Map of the 6.7\,GHz methanol masers toward NIRS3. Triangles and circles represent maser spots before and after the burst, respectively. The velocity-integrated emission of the 6.7\,GHz masers (gray scale) and the  JVLA 5\,GHz radio continuum emission(red contours) are also shown.}
    \label{fig1}
\end{wrapfigure}
Thus, such events are likely accompanied by flares of Class II methanol masers as a consequence of the IR pumping theory. For this reason, we initiated NIR imaging of the massive star forming region S255IR two weeks after the flare of the associated 6.7\,GHz methanol maser was reported (\cite{fujisawa_atel_2015}). It revealed the outburst of NIRS3, a $\sim$20\,M$_\odot$ MYSO (SIMBAD designation
\href{http://simbad.u-strasbg.fr/simbad/sim-id?Ident=[TGJ91]+S255+NIRS+3&NbIdent=1&Radius=2&Radius.unit=arcmin&submit=submit+id}{([TGJ91] S255 NIRS 3}, \cite{stecklum_methanol_2016}). Our extensive observing campaign yielded major properties of the HMYSO burst (\cite{caratti_o_garatti_disk-mediated_2016}). 
These results confirmed that OB stars may form via disk accretion and that their disks are prone to instabilities leading to accretion bursts, which trigger flares of Class II methanol masers. At about the same time as the  burst of NIRS3 went off, a similar event was detected in the submm/mm which showed maser flares as well (\cite{hunter_extraordinary_2017}, this volume). This provided independent support for the conclusions drawn above.


For what concerns the temporal behavior of the methanol masers during the burst of NIRS3, it was found that the 5.9\,km/s component started to fade $\sim$15\,d after reaching peak level while a new feature at 6.5\,km/s emerged which reached its peak flux $\sim$20\,d later than the 5.9\,km/s component (K. Fujisawa, priv. comm.). Our EVN and  JVLA observations after the burst revealed that the main pre-burst maser cluster vanished and a new, extended region of 6.7\,GHz maser emission surfaced further out (500$\dots$1000\,au) (\cite{mosca_17}, Fig.\ref{fig1}). From the delay between the rise of the old and new maser components and their projected separation a lower limit to the propagation speed of the excitation
could be derived for the first time which amounts to ${\sim}0.15c$. There are several possible reasons for the
 subluminal velocity. Since the energy transport by photons rests on absorption and re-emissionas well as scattering, it will become slower at increasing optical depth. The heating of dust beyond the snow line will be delayed by the endothermic sublimation of frozen volatiles. Dust growth in the dense YSO environment will have a similar effect due to the increase in grain heat capacity. 

Notably, the radio continuum stayed constant during the burst and eventually started to rise $\sim$300\,d after the flare detection (Cesaroni ea., A\&A subm.).


\section{The first IR light-curve for a periodic maser source}

The lack of complementary data, in particular time-resolved IR photometry, hinders to disclose the driving mechanisms for periodic masers. The IR pumping of Class II methanol masers suggests that cyclic luminosity variations of YSOs  will cause maser periodicity. In order to confirm this claim, we tried to establish IR light curves of periodic maser sources from IRAC and NEOWISE photometry. No successful ground-based attempts in this respect were reported  so far. Unfortunately,  useful data is very scarce since luminous YSOs are generally saturated in space-based IR imaging. This is not the case for G107.298+5.639 (G107 for short, SIMBAD designation
 \href{http://simbad.u-strasbg.fr/simbad/sim-id?Ident=[PFG2013]+I22198-MM2-S&NbIdent=1&Radius=2&Radius.unit=arcmin&submit=submit+id}{[PFG2013] I22198-MM2-S}), a less luminous, intermediate-mass YSO (\cite{sanmong_2010}), situated at $750\pm27$\,pc (\cite{hiro_2008}), which nevertheless excites periodic methanol and water masers (\cite{fujisawa_periodic_2014,szymczak_discovery_2016}).
\begin{figure*}[h]
    \centering
    \includegraphics[width=4.025cm, angle=0]{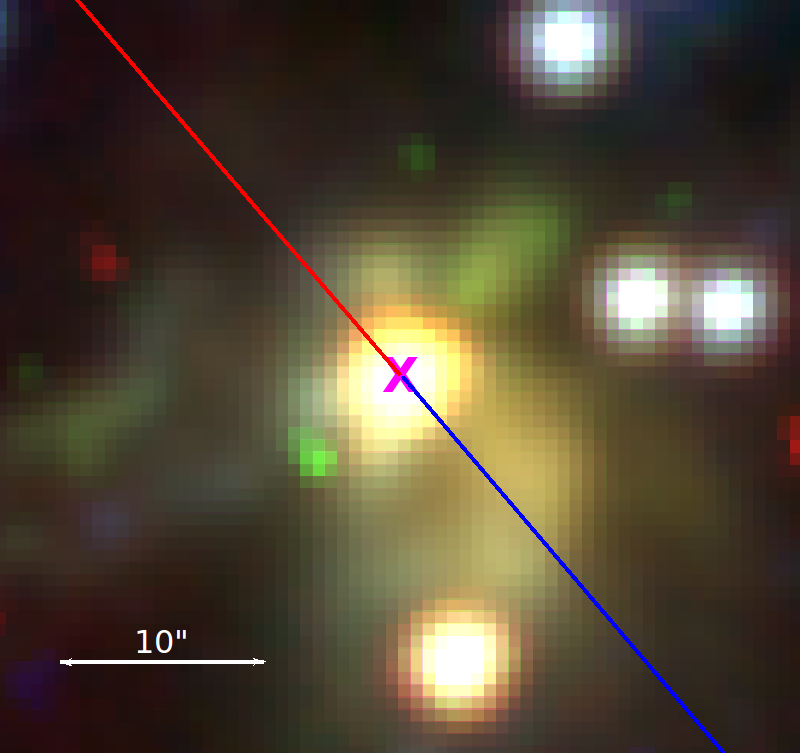}
    \includegraphics[width=7.5cm, angle=0]{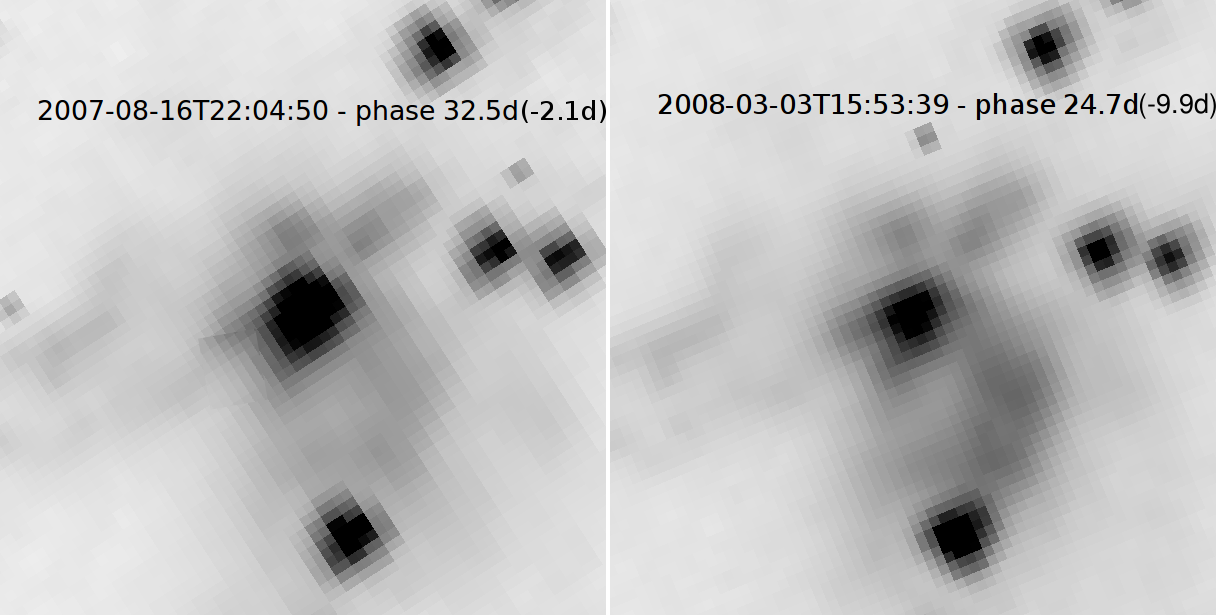}
    \caption{{\bf Left} -  IRAC RGB-color image (channels 3, 2, 1) of G107 (cross). The line indicates the rotation axis of the circumstellar disk
and the sense of the outflow (red-/blue-shifted). 
{\bf Center/right} -  4.5\,$\mu$m images for the two epochs at phases of 32.5 and 24.7\,d, respectively. The out of phase  variability of the YSO and the nebulosity is obvious.
}
\label{fig2}
\end{figure*}

Spitzer/IRAC observations at two epochs (PI G. Fazio, Fig.\,\ref{fig2}) clearly reveal variability of both the YSO and the nebulosity associated with the blue-shifted outflow lobe. The out of phase brightness change of the latter indicates a light echo.

\begin{figure}[t]
\begin{center}
    \includegraphics[width=9cm]{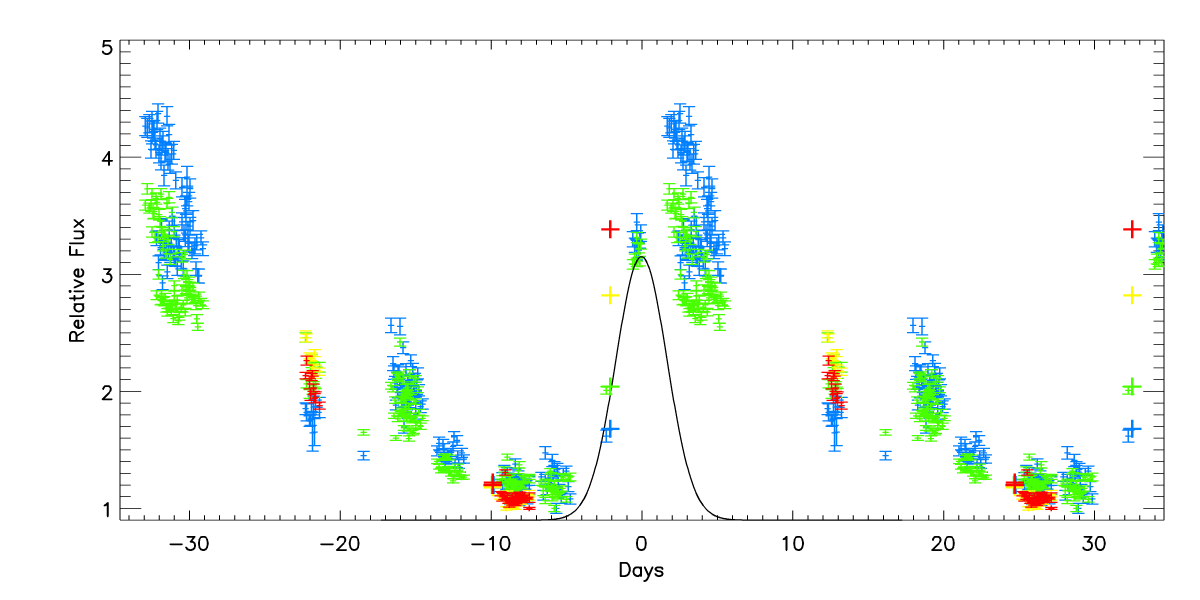}
    \vspace*{-3mm}
    \caption{NEOWISE multi-color ({\color{blue} W1},{\color{green} W2}, {\color{yellow} W3}, {\color{red} W4}) light curve of G107, folded by the 34.6\,d period, and normalized to minimum brightness. Zero represents the date of the maser flare peak. Large crosses mark IRAC flux ratios, normalized to the low state. The Gaussian fit to the methanol maser light curve (\cite{fujisawa_periodic_2014}) is shown for comparison. The apparent jumps at phase of $\sim$4\,d are due to long-term brightening.}
    \label{fig3}
 \end{center}
\end{figure}

From NEOWISE data, we established the light curve of G107 (Fig.\,\ref{fig3}) which has 
the same period  
as that of the masers (34.6\,d, \cite{fujisawa_periodic_2014}). However, its saw tooth shape is 
distinct from that of the maser flares, and resembles light curves of Cepheids. While the 3.6\,$\mu$m and 4.5\,$\mu$m IRAC fluxes
of the two epochs preceding WISE are weaker, their ratios 
fit the normalized WISE light curve. This points to a quite stable period over $\sim$10 years as well as long-term brightening. Notably,  G107 does not fit the period-luminosity relation for pulsating massive YSOs (\cite{inayoshi_direct_2013}). Moreover, the skewness of the light curve contradicts the eccentric binary accretion model (\cite{art_lub_1996}). So the question on the driving mechanism for the periodicity is still open.

\vspace*{-.25cm}
\section{Conclusions}
The change of the maser population of NIRS3 due to its burst
won't be permanent. Since the burst ceased, they will redistribute according to the cooling of the circumstellar environment. This provides a unique opportunity to study the dynamics of maser excitation. The synergy between radio and IR observations is essential to disclose the reasons of maser variability.  While the ongoing NEOWISE mission as well as the VVV(X) survey are extremely helpful in this respect, their cadence is not sufficient. Dedicated IR imaging capability is required to follow-up maser flares in a target-of-opportunity fashion.

\vspace*{-.25cm}

\end{document}